\documentclass[journal]{IEEEtran}
\usepackage{graphicx}
\usepackage{amsmath}
\usepackage{cite}
\usepackage{subfigure}
\usepackage{bm}
\ifCLASSINFOpdf
\else
\fi
\begin{document}

\title{Turbulence-Resistant All Optical Relaying Based on Few-Mode EDFA in Free-Space Optical Systems}

%

\author{Shanyong~Cai,~\IEEEmembership{Member,~OSA,}
        Zhiguo~Zhang,
        and~Xue~Chen
\thanks{Shanyong Cai, Zhiguo Zhang and Xue Chen are with State Key Laboratory of Information Photonics and Optical Communications, Beijing University of Posts and Telecommunications, Beijing, 100876, China. e-mail: (caishanyong@bupt.edu.cn, zhangzhiguo@bupt.edu.cn, xuechen@bupt.edu.cn).}

\thanks{Manuscript received xxx xx, xxxx; revised xxx xx, xxxx. This work is supported by National Natural Science Foundation of China (No. 61671076 and No. 61801043) and ``the Fundamental Research Funds for the Central Universities''.}}

%
%

\markboth{Journal of \LaTeX\ Class Files,~Vol.~xx, No.~xx, xxx~xxxx}%
{Shell \MakeLowercase{\textit{et al.}}: Bare Demo of IEEEtran.cls for IEEE Journals}
%



\maketitle

\begin{abstract}
We investigate the communication performance of a few-mode EDFA based all-optical relaying system for atmospheric channels in this paper. A dual-hop free space optical communication model based on the relay with two-mode EDFA is derived. The BER performance is numerically calculated. Compared with all-optical relaying system with single-mode EDFA, the power budget is increased by $\text{4 dB}$, $\text{7.5 dB}$ and $\text{11.5 dB}$ at BER = 1$\bm{\times}$10$^{\text{-4}}$ under the refractive index structure constant C$_\text{n}^\text{2}$ = 2$\bm{\times}$10$^{\text{-14}}$, 5$\bm{\times}$10$^{\text{-14}}$ and 1$\bm{\times}$10$^{\text{-13}}$ respectively when a few mode fiber supporting 4 modes is utilized as the receiving fiber at the destination. The optimal relay location  is slightly backward from the middle of the link. The BER performance is the best when mode-dependent gain of FM-EDFA is zero.

\end{abstract}

\begin{IEEEkeywords}
free space optical communication, relay, few mode fiber, EDFA.
\end{IEEEkeywords}

\IEEEpeerreviewmaketitle

\section{Introduction}

\IEEEPARstart{R}{ecently}, free-space optical (FSO) communication systems have attracted a lot of attention for its high bandwidth, unlicensed spectrum and high security \cite{khalighi-icst-2014}. However, in FSO system, atmospheric turbulence results in reduced transmission distance and degraded bit error rate (BER). A variety of techniques have been proposed to mitigate atmospheric turbulence fading including spatial diversity \cite{zhu-IEEtc-2002, navidpour-IEEETWC-2007}, wavefront correction \cite{love-ao-1997} and relay-assisted optical communication system \cite{safari-IEEETWC-2008,karimi-jlt-2009,rjeily-IEEETC-2011,karimi-jlt-2011,kazemlou-jlt-2011,bayaki-IEEETC-2012,kashani-ICL-2012,nor-jlt-2017,kashani-jocn-2013,ruiz-jocn-2018}. Relay-assisted transmission system relays the data signal from the source to destination through intermediate nodes (relays). It exploits the fact that the turbulence induced fading variance is distance-dependent and yields improved transmission distance and BER by taking advantage of the shorter hop distances \cite{kashani-jocn-2013}. Moreover, relaying systems support optical connection between source and destination which do not have a line of sight.

The current literature on relaying system including serial and parallel electrical relaying and all-optical relaying \cite{safari-IEEETWC-2008,karimi-jlt-2011}. Compared with electrical relaying where the relays employ optical-to-electrical (OE) and electrical-to-optical (EO) convertors, all-optical relaying avoids the requirement for high-speed electronics and electro-optics. The state-of-art all-optical relaying system is based on single-mode (SM) erbium-doped fiber amplification (EDFA) which only receives and amplifies the power of spatial fundamental mode (Gaussian) \cite{kazemlou-jlt-2011,bayaki-IEEETC-2012,kashani-ICL-2012,nor-jlt-2017}. Thus, the turbulence induced fading variance at the relay is high which limits the hop distance. In fact, the distorted wavefront after atmospheric channel is a superposition of spatial fundamental mode and spatial high-order modes (e.g. Laguerre-Gaussian). If multiple spatial modes are received and amplified, the turbulence induced fading at the relay will be improved. Recently, spatial-division multiplexing (SDM) optical fiber communication system has attracted a lot of attention \cite{richardson-nature-2013,li-aop-2014} for its ability to increase system capacity. Few mode fiber (FMF) components like FMFs \cite{sillard-ofc-2016} and few-mode erbium-doped fiber amplifications (FM-EDFAs) \cite{bai-oe-2012} have been fabricated and implemented in SDM optical fiber communication system. FMF components have also been paid attention to in FSO systems \cite{ozdur-ol-2013,arikawa-ecoc-2016,zheng-oe-2016,huang-ecoc-2017,geisler-spie-2018,zheng-oe-2018}. For example, experimental results show that the relative standard deviation (RSD) of turbulence fading can be significantly improved by using a FMF receiver compared with single mode fiber (SMF) receiver \cite{zheng-oe-2016}. Moreover, FM-EDFAs which could receive and amplify several spatial modes simultaneously have been used in preamplified receiver in FSO system at ECOC'17 to resist turbulence \cite{huang-ecoc-2017}.

In this paper, a scheme of FM-EDFA based all-optical relaying system to resist turbulence for atmospheric channels is proposed which is the extension of our earlier work on CLEO'18 \cite{cai-cleo-2018}. The relay is composed of a two-mode EDFA with two mode fiber (TMF) pigtails. At the relay, the TMF pigtail is utilized as a receiver leading to the improvement of the turbulence induced fading. After FM-EDFA, the mixed spatial modes field is equally amplified and forward to the destination. Compared with previous SM-EDFA based relaying system, the simulation resuls show that system BER is optimized for both fixed and variable gain amplification under atmospheric turbulence.

\section{DUAL-HOP FSO MODEL BASED ON FM-EDFA}
In this section, we develop model for dual-hop FSO system based on FM-EDFA. Firstly, the scheme is given. Secondly, the signals and noises at the destination terminal for dual-hop FSO communication systems are obtained according to the general approach introduced in [10].

\begin{figure*}[htb]
\centering
\includegraphics[width=15cm]{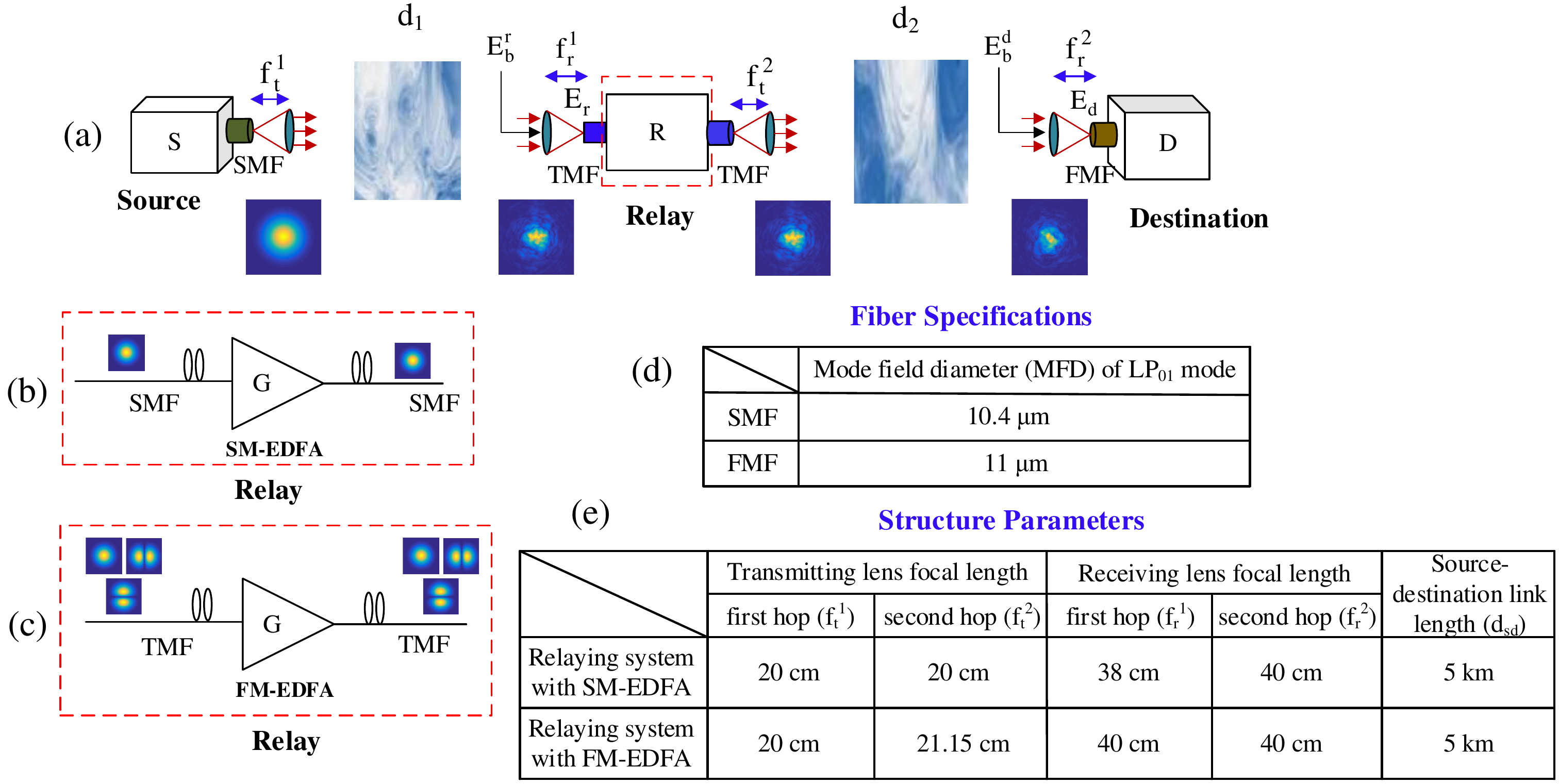}
\caption{ The schematic of (a) dual-hop all-optical relaying system based on FM-EDFA, (b) the relay with SM-EDFA, and (c) the relay with FM-EDFA. (d) SMF/FMF's specifications. (e) Structure parameters for all-optical relaying system with SM-EDFA/FM-EDFA.}
\end{figure*}

\subsection{Scheme }
The scheme of FM-EDFA based dual-hop transmission system is shown in Fig. 1(a). The relay is composed of receiving lens, FM-EDFA and transmitting lens (Fig. 1(a,c)). At the input end of relay, the signal and background light is coupled into FMF pigtail of FM-EDFA. In the remainder of this paper, assuming FM-EDFA only supports two modes (LP$_{\text{01}}$ mode and degenerate LP$_{\text{11}}$ modes). Assuming the mode dependent gain of FM-EDFA is zero, the noisy spatial mode fields are equally amplified by FM-EDFA and forwarded to the destination. A FMF which supports N modes (N=1, 2 or 4) is placed on the destination to collect light field. The optical links between source and relay and between relay and destination are atmospheric turbulence channels. The light field is distorted by atmospheric turbulence in the links.

The fiber specifications and link parameters are shown in Fig. 1(d-e). The source-destination link length is $5.0 km$. The relay is placed in the middle of the link ($d_1 = d_2$). SMF at Source/Relay or FMF at Relay/Destination is placed on the focal point of transmiting/receiving lens. SMF is commercially available SMF-28 fiber. The mode field diameter of SMF-28 is $10.4 \mu m$  (Fig. 1(d)). FMF is taken to be graded-index fiber with profile parameter $\rho=2$, thus the spatial distribution of core mode in the FMF is given by
\begin{equation}
\begin{split}
{\psi_{{m},{n}}}\left( {r,\theta } \right) = C*L_{n - 1}^m\left( {\frac{{{r^2}}}{{{\omega ^2}}}} \right)*{\left( {\frac{r}{\omega }} \right)^m}*{\rm{exp}}( - \frac{{{r^2}}}{{2{\omega ^2}}})\\
*\left\{ {\begin{array}{*{20}{c}}
{cos(m\theta )}\\
{sin(m\theta )}
\end{array}} \right.,
\end{split}
\end{equation}
where $C$ is the amplitude constant, $L_n^m$ are the associated Laguerre polynomials, and $\omega$ is a constant relating modal diameter. Here, $\omega = 3.89 \mu m$ is set, so the mode field diameter of LP$_{\text{01}}$ mode is $11 \mu m$ which is equal to that of OFS' FMFs  (Fig. 1(d)). For relaying system with FM-EDFA, the transmitting lens focal lengths of the first and second hop are $f_t^1 = 20 cm$ and $f_t^2 = 21.15 cm$ respectively (Fig. 1(a,e)) to ensure that the diameter of the output beams are the same. The receiving lens focal lengths of the first and second hop are equal to 40 cm ($f_r^1 = f_r^2 =40 cm$) (Fig. 1(e)). In addition, in relaying system with SM-EDFA, the receiving lens focal lengths of the first hop is set to 38 cm for fair comparison with relaying system with FM-EDFA (Fig. 1(e)). At this time, the coupling efficiencies from free space to LP$_{\text{01}}$ mode in SMF/FMF are the same and all equal to 44\% without atmospheric turbulence.

Compared with SM-EDFA based all-optical relay (Fig. 1(b)), FM-EDFA based relay characterizes lower turbulence fading RSD. In analogy with aperture averaging, the reason can be attributed to mode averaging where multiple spatial modes contained in the turbulence induced distorted light field are received simultaneously. In addition, the average receiving power at the relay with FM-EDFA is higher than that with SM-EDFA. In moderate to strong turbulence, the average receiving power can be increased by 4 dB for TMF compared with SMF \cite{zheng-oe-2016}. After SM-EDFA or FM-EDFA, the fixed amplification gain of EDFA keeps the average relay output power constant at $P_r$.

\subsection{Model}

\subsubsection{Channel Model}

The considered system operates at wavelength $\lambda = 1550 nm$ with on-off keying (OOK) and direct detection (DD). The temporal random atmospheric channel impairments are denoted by $h_i(t)$ for the $i$th hop. The channel impairments $h_i(t)$ usually can be formulated as $h_i(t)\approx \sqrt{L_i} h_p^i\eta_i$. $L_i = \alpha_{attn}d_i$ denotes the atmospheric attenuation caused by absorption and scattering effects where $\alpha_{attn}$ and $d_i$ represent the attenuation coefficient and hop distance, respectively. It can be considered constant since the weather changes slowly over a long time. $h_p^i$ denotes impairment caused by the spatial spread of the optical beam and also can be considered constant for a given transmitting beam diameter and link distance. $\eta_i$ denotes turbulence-induced fading factor and is a random coefficients. In this paper, to obtain $h_i$ when FMF is utilized as receiving fiber in the two hops, atmospheric turbulence is simulated by phase plates with appropriate randomness.

Turbulence phase plates are generated via Fourier Transform method \cite{xu-cleo-2014}. The basic idea is using atmospheric phase spectrum according to atmospheric turbulence theory to filter Gaussian random noises. The phase screen in x-space is calculated by
\begin{equation}
\varphi(x,y)=IFFT(C\cdot\sigma(k_x,k_y))
\end{equation}
\begin{equation}
\sigma^2(k_x,k_y)=0.023r_0^{-\frac{5}{3}}[k_x^2+k_y^2]^{-\frac{11}{6}},
\end{equation}
where $\sigma^2(k_x,k_y)$ is the variance of phase spectrum, and $C$ is a matrix of complex Gaussian random numbers. $r_0$ is the atmospheric coherence length, which denotes the turbulence strength and is a function of refractive index structure constant, link distance and optical beam. For Gaussian beam, $r_0$ can be approximated by the simple expression \cite{andrews-2005}
\begin{equation}
r_0 = [\frac{8}{3(a+0.62\Lambda^{11/6})}]^{3/5}(0.423C_n^2k^2d)^{-3/5},    l_0 \ll r_0 \ll L_0,
\end{equation}
where $C_n^2$ is the refractive index structure constant, $k$ is the wave number, $d$ is link distance, $a$ and $\Lambda$ are the parameters related to the beam waist, divergence angle of transmitting Gaussian beam and transmission distance ($d$). The detailed derivation of $a$ and $\Lambda$ can be referred to reference \cite{andrews-2005}. In this paper, $D/r_0$ is used to characterize the effect of turbulence on FSO beams where $D=38 mm$ is the transmitting beam diameter in our scheme. Source-destination length $d_{sd}$ is used to calculate $D/r_0$.

\begin{figure}[htb]
\centering
\includegraphics[width=7.5cm]{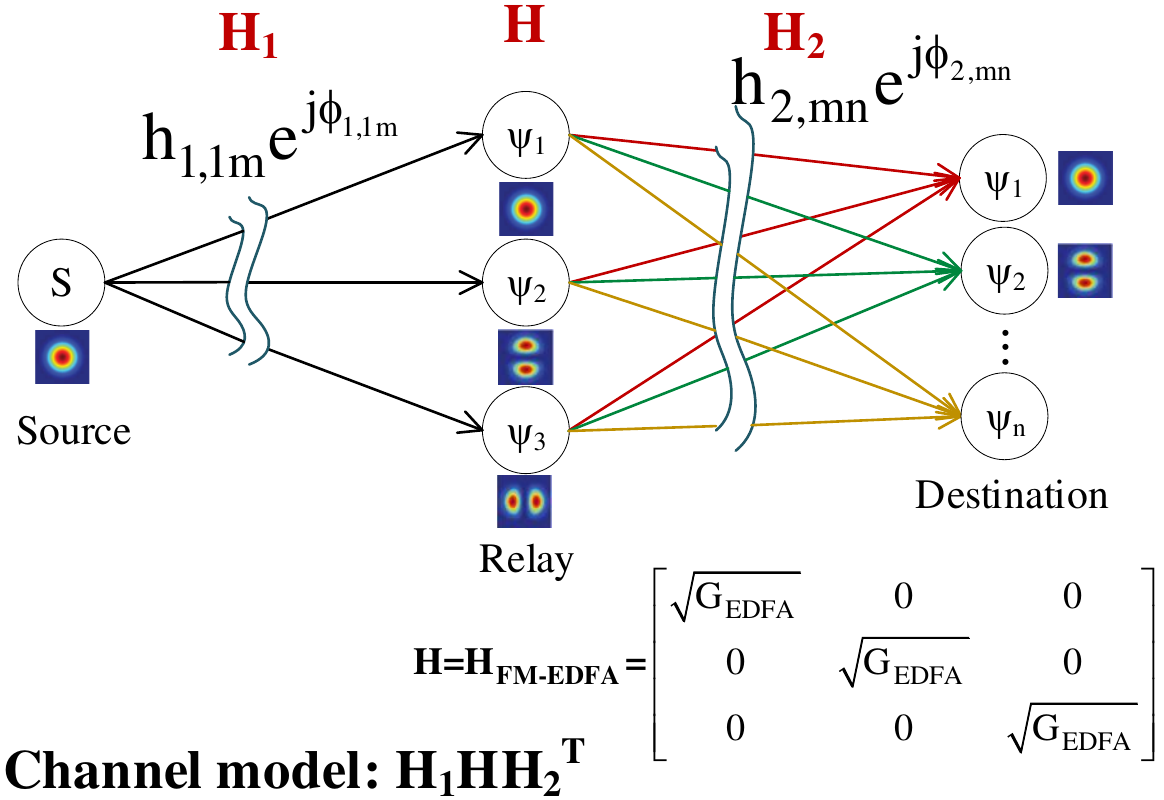}
\caption{Channel model of the dual-hop relaying system with FM-EDFA.}
\end{figure}

\subsubsection{Dual-Hop Transmission Model Based on FM-EDFA}

A diagram of dual-hop model with FM-EDFA is shown in Fig. 2. For relay, at the input port of FM-EDFA, the total received electric field is composed of three spatial modes (including two degenerate modes) and is the sum of the signal electric field and the background radiation electric field
\begin{equation}
\begin{aligned}
E_r(r,\theta,t)=\sqrt{2xP_t}\sum_{m=1}^{3}\Re\{h_{1,1m}e^{j\phi_{1,1m}}e^{j\omega_0t}
\psi_{m}(r,\theta)\}\\ +\sqrt{2N_b\delta\nu}\sum_{m=1}^{3}\sum_{l=-M}^{M}\Re\{e^{j\phi_{l,m}}e^{j\omega_lt}\psi_{m}(r,\theta)\},
\end{aligned}
\end{equation}
where $P_t$ and $x\in\{0,2\}$ denote the transmit power of the source and the transmitted OOK symbol. $h_{1,1m}e^{j\phi_{1,1m}}$ represent fading coefficients from transmitting LP$_{\text{01}}$ mode to the $m$th fiber mode of FM-EDFA in the first hop ($m=1$ corresponds to LP$_{\text{01}}$, $m=2$ corresponds to LP$_{\text{11a}}$ and $m=3$ corresponds to LP$_{\text{11b}}$, respectively.). $N_b = P_b/B_o$ denotes the background radiation power spectral density, where $P_b$ and $B_o$ are the background radiation power of single  mode and the optical bandwidth of the system, respectively. The background radiation noise is modeled as a white Gaussian noise in space, thus, values of $N_b$ corresponding to three spatial modes are the same. Following the approach in \cite{olsson-jlt-1989}, the electric field corresponding to the background radiation in (5) is written as a sum of $2M+1$ ($M = B_o/(2\delta\nu)$) cosine terms at frequencies $\omega_l = \omega_0 +2\pi l\delta\nu$ with center frequency $\omega_0 =2\pi\nu_0$ ($\nu_0$: optical center frequency), where $\delta\nu$ denotes the spacing of the considered frequencies. $\phi_{l,m}$ denotes a random phase corresponding to the $m$th mode. $\psi_{m}(r,\theta)$ is the spatial complex amplitude of the $m$th fiber mode.

FM-EDFA is followed to amplify the received electric field. ASE noise is introduced after FM-EDFA. For amplification, a fixed gain or a variable gain can be employed. The fixed amplification gain keeps the average output power of the relay limited at $P_r$ and is expressed as:
\begin{equation}
\begin{aligned}
G_{fixed}=\frac{P_r+3n_{sp}h\nu_0B_o}{P_tE[\alpha_1]+3P_b+3n_{sp}h\nu_0B_o},
\end{aligned}
\end{equation}
where $E[\alpha_1]$ is the average channel impairment of first hop with $\alpha_1 = \sum_{m=1}^{3}h_{1,1m}^2$. $3P_b$ represents the total background radiation power of three modes. $n_{sp}$ is the amplifier spontaneous emission factor. $h$ denotes Planck`s constant. The variable amplification gain keeps the output power of relay stabilized at constant value $P_r$ at all times and is given by
\begin{equation}
\begin{aligned}
G_{var}=\frac{P_r+3n_{sp}h\nu_0B_o}{P_t\alpha_1+3P_b+3n_{sp}h\nu_0B_o}.
\end{aligned}
\end{equation}
Channel impairment $\alpha_1$, required in the above equation can be estimated at the relay based on pilot symbols emitted by the source. For the sake of brevity, in the remainder of this paper, $G_{EDFA}$ represents both $G_{fixed}$ and $G_{var}$ of the relay  and supposing $P_r = P_t$.

The total electric field at the destination is given by
\begin{equation}
E_d(r,\theta,t) = \sum_{n}E_{d,n}(t)\psi_{n}(r,\theta),
\end{equation}
which is the superposition of multiple spatial mode fields ($\psi_{n}(r,\theta)$) supported by the receiving FMF in the destination. The time dependent component $E_{d,n}(t)$ is given by
\begin{equation}
\begin{aligned}
E_{d,n}(t) = E_s(t) + E_{b,r}(t) + E_{ASE}(t) + E_{b,d}(t)\\ = \sqrt{2xP_t}|\textbf{H$_1$HH$_2^T$}|\cos(\omega_0t + \phi_{s}) \\ + \sqrt{2N_b\delta\nu G_{EDFA}}\sum_{m=1}^{3}\sum_{l=-M}^{M}h_{2,mn}\cos(\omega_lt + \phi_{r,l,mn})\\ + \sqrt{2N_0\delta\nu}\sum_{m=1}^{3}\sum_{l=-M}^{M}h_{2,mn}\cos(\omega_lt + \phi_{ASE,l,mn})\\+  \sqrt{2N_b\delta\nu}\sum_{l=-M}^{M}\cos(\omega_lt + \phi_{d,l,n}),
\end{aligned}
\end{equation}
which is the add of signal light electric field $E_s$, background radiation electric field at the relay $E_{b,r}$, ASE noise after FM-EDFA $E_{ASE}$ and background radiation electric field at the destination $E_{b,d}$. $\textbf{H$_1$} = [h_{1,11}e^{j\phi_{1,11}}, h_{1,12}e^{j\phi_{1,12}}, h_{1,13}e^{j\phi_{1,13}}]$, $\textbf{H$_2$}=[h_{2,1n}e^{j\phi_{2,1n}}, h_{2,2n}e^{j\phi_{2,2n}}, h_{2,3n}e^{j\phi_{2,3n}}]$ and $\textbf{H$_1$HH$_2^T$}=\sum_{m=1}^{3}\Re\{h_{1,1m}e^{j\phi_{1,1m}}\sqrt{G_{EDFA}}h_{2,mn}e^{j\phi_{2,mn}}\}$ where
$h_{2,mn}e^{j\phi_{2,mn}}$ denote the fading coefficients from the $m$th spatial mode at relay to the $n$th spatial mode at the destination for the second hop (Fig. 2). $N_0=n_{sp}(G_{EDFA}-1)h\nu_{0}$ is the spectral density of the ASE noise. $\phi_{r,l,mn}$, $\phi_{ASE,l,mn}$ and $\phi_{d,l,n}$ in (9) are mutual independent random phase. At the destination, after photodetector (PD), the optical signal is converted into electrical signal. The expression of the photocurrent is as follows:
 \begin{equation}
\begin{aligned}
i_{photo}(t)=R\sum_{n}E_{d,n}^2(t),
\end{aligned}
\end{equation}
where $R=\frac{\rho q}{h\nu_0}$ denotes the responsivity of the PD, $\rho$ and $q=1.6\times 10^{-19}$ denote the efficiency of PD and the charge of an electron, respectively. The useful signal terms and unwanted noise terms can be obtained by inserting (9) into (10). Following the approach in reference \cite{bayaki-IEEETC-2012}, the expressions of these terms are as follows.\\
\textit{Signal Direct Current term :}
\begin{equation}
\begin{aligned}
I_s = \sum_{n}RxP_t|\textbf{H$_1$HH$_2^T$}|^2.
\end{aligned}
\end{equation}\\
\textit{Background-Background Direct Current terms :}
\begin{equation}
\begin{aligned}
I_{b\times b,r} = \sum_{n}RN_bB_oG_{EDFA}\sum_{m=1}^{3}h_{2,mn}^2,
\end{aligned}
\end{equation}

\begin{equation}
\begin{aligned}
I_{b\times b,d} = \sum_{n}RN_bB_o,
\end{aligned}
\end{equation}
where $I_{b\times b,r}$ and $I_{b\times b,d}$ denote the background-background direct current at relay and destination, respectively.\\
\textit{Background-Background Beat Noise:}
\begin{equation}
\begin{aligned}
\sigma_{b\times b,r}^2 = \sum_{n}R^2N_b^2G_{EDFA}^2(2B_eB_o-B_e^2)\\(\sum_{m=1}^{3}h_{2,mn}^4 + 4\sum_{i=1}^{3}\sum_{j=1,j\neq i}^{3}h_{2,in}^2h_{2,jn}^2),
\end{aligned}
\end{equation}

\begin{equation}
\begin{aligned}
\sigma_{b\times b,d}^2 = \sum_{n}R^2N_b^2(2B_eB_o-B_e^2),
\end{aligned}
\end{equation}

\begin{equation}
\begin{aligned}
\sigma_{b,r\times d}^2 = \sum_{n}4R^2N_b^2G_{EDFA}(\sum_{m=1}^{3}h_{2,mn}^2)(2B_eB_o-B_e^2),
\end{aligned}
\end{equation}
where $\sigma_{b\times b, r}^2$, $\sigma_{b\times b, d}^2$ and $\sigma_{b,r\times d}^2$ denote the variance of the relay background-background radiation beat noise, the destination background-background radiation beat noise and the relay-destination background beat noise, respectively. $B_e$ is the electronic bandwidth of the system. The total background-background radiation beat noise variance is given by
\begin{equation}
\begin{aligned}
\sigma_{b\times b}^2 = \sigma_{b\times b, r}^2+\sigma_{b\times b, d}^2+\sigma_{b,r\times d}^2.
\end{aligned}
\end{equation}\\
\textit{ASE-ASE Direct Current term :}
\begin{equation}
\begin{aligned}
I_{ASE\times ASE} = \sum_{n}RN_0B_o(\sum_{m=1}^{3}h_{2,mn}^2).
\end{aligned}
\end{equation}\\
\textit{ASE-ASE Beat Noise:}
\begin{equation}
\begin{aligned}
\sigma_{ASE\times ASE}^2 = \sum_{n}R^2N_0^2(2B_eB_o-B_e^2)\\(\sum_{m=1}^{3}h_{2,mn}^4 + 4\sum_{i=1}^{3}\sum_{j=1,j\neq i}^{3}h_{2,in}^2h_{2,jn}^2).
\end{aligned}
\end{equation}\\
\textit{Signal-Background Beat Noise:}
\begin{equation}
\begin{aligned}
\sigma_{s\times b,r}^2 = \sum_{n} 4R^2xP_t|\textbf{H$_1$HH$_2^T$}|^2G_{EDFA}(\sum_{m=1}^{3}h_{2,mn}^2)N_bB_e,
\end{aligned}
\end{equation}

\begin{equation}
\begin{aligned}
\sigma_{s\times b,d}^2 = \sum_{n} 4R^2xP_t|\textbf{H$_1$HH$_2^T$}|^2N_bB_e,
\end{aligned}
\end{equation}
where $\sigma_{s\times b,r}^2$ and $\sigma_{s\times b,d}^2$ denote the variance of beat noise between signal and relay background radiation and that of beat noise between signal and destination background radiation, respectively. \\
\textit{Signal-ASE Beat Noise:}
\begin{equation}
\begin{aligned}
\sigma_{s\times ASE}^2 = \sum_{n} 4R^2xP_t|\textbf{H$_1$HH$_2^T$}|^2(\sum_{m=1}^{3}h_{2,mn}^2)N_0B_e.
\end{aligned}
\end{equation}\\
\textit{Background-ASE Beat Noise:}
\begin{equation}
\begin{aligned}
\sigma_{b\times ASE}^2 = \sum_{n} 4R^2N_bN_0(2B_oB_e-B_e^2)(\sum_{m=1}^{3}h_{2,mn}^2)\\(1+G_{EDFA}\sum_{m=1}^{3}h_{2,mn}^2),
\end{aligned}
\end{equation}
which is composed of relay background-ASE beat noise and destination background-ASE beat noise.\\
\textit{Shot Noise:}
\begin{equation}
\begin{aligned}
\sigma_{shot, off}^2 = 2q(I_{b\times b,r}+I_{b\times b,d}+I_{ASE\times ASE})B_e,
\end{aligned}
\end{equation}
\begin{equation}
\begin{aligned}
\sigma_{shot, on}^2 = \sigma_{shot, off}^2 + 2qI_sB_e,
\end{aligned}
\end{equation}
where $\sigma_{shot, off}^2$ and $\sigma_{shot, on}^2$ denote the shot noise variances for transmit symbol $x=0$ and $x=2$, respectively.

In addition to the noise terms discussed above, the impact of thermal noise of PD at the destination is also important. Finally, the total noise variance of relaying system is
\begin{equation}
\begin{aligned}
\sigma_{off}^2 = \sigma_{th}^2 + \sigma_{shot, off}^2 + \sigma_{b\times b}^2 +\sigma_{ASE\times ASE}^2 +\sigma_{b\times ASE}^2,
\end{aligned}
\end{equation}
for transmit symbol $x=0$, and
\begin{equation}
\begin{aligned}
\sigma_{on}^2 = \sigma_{off}^2 + 2qI_sB_e + \sigma_{s\times b,r}^2 + \sigma_{s\times b,d}^2 + \sigma_{s\times ASE}^2,
\end{aligned}
\end{equation}
for transmit symbol $x=2$ with thermal noise variance $\sigma_{th}^2 = 4KTB_e/R_L$. $K$ is the Boltzmann constant, $T$ is the temperature in Kelvin, and $R_L$ is the photodetector load resistance. In general, the model for dual-hop transmission system based on FM-EDFA is composed of signal current $I_s$, in (11) and the noise variances in (26) and (27). In the remainder of this paper, the performance comparison between dual-hop transmission system with FM-EDFA and dual-hop transmission system with SM-EDFA will be given where the model for dual-hop transmission system based on SM-EDFA can be obtained when only LP$_{\text{01}}$ mode is considered at the relay.

\section{NUMERICAL RESULTS}
To evaluate the system performance, we present simulation results for BER of dual-hop all-optical relaying system with FM-EDFA. At first, the system parameters of the relaying system is given in Table 1. The system operates at bit-rates (BR) 2 Gpbs. The BER is averaged over 1000 different fading states with $2 \times 10^6$ bits transmitted per fading state.

\begin{table}[htb]
\caption{System Parameters}
\centering
\begin{tabular}{|c|c|}
\hline
Parameter & value \\
\hline
Wavelength ($\lambda$) & $1550~nm$ \\
\hline
Electrical bandwidth ($B_e$) & $2~GHz$\\
\hline
Optical bandwidth ($B_o$) & $125~GHz$\\
\hline
Data rate ($R_b$) & $2~Gbps$\\
\hline
Amplifier spontaneous emission factor ($n_{sp}$) & $1.4$ \\
\hline
Receiver noise temperature ($T$) & $300~K$\\
\hline
Receiver quantum efficiency ($\rho$) & $0.75$\\
\hline
Photodetector load resistance ($R_L$) & $50~\Omega$\\
\hline
Background radiation energy ($P_b$) & $20~nW$\\
\hline
Photodetector load resistance ($R_L$) & $50~\Omega$\\
\hline
\end{tabular}
\end{table}

In the presence of atmospheric turbulence, atmospheric fading coefficients $h_i$ are time-varying. With the aid of the 1000 turbulence screens generated by Fourier Transform method, the 1000 values of $h_i$ are calculated over $N=1000$ fading states. Considering the refractive index structure constant $C_n^2=1\times 10^{-13}$ and no atmospheric attenuation, the statistical distribution of $1000$ values of $\alpha_1$ ($\alpha_1 = \sum_{m=1}^{3}h_{1,1m}^2$ denotes the atmospheric fading at the relay) is displayed in Fig. 3. Obviously, the average value of fading at the relay with FM-EDFA ($E(\alpha_{1, FM})=9.84 dB$) is smaller than that of the relay with SM-EDFA ($E(\alpha_{1, SM})=12.5 dB$). The atmospheric fading at the relay with FM-EDFA shows a narrower distribution (RSD is restrained by 69\%) which is expected to bring improved BER for all optical relaying system.

\begin{figure}[htb]
\centering
\includegraphics[width=7.5cm]{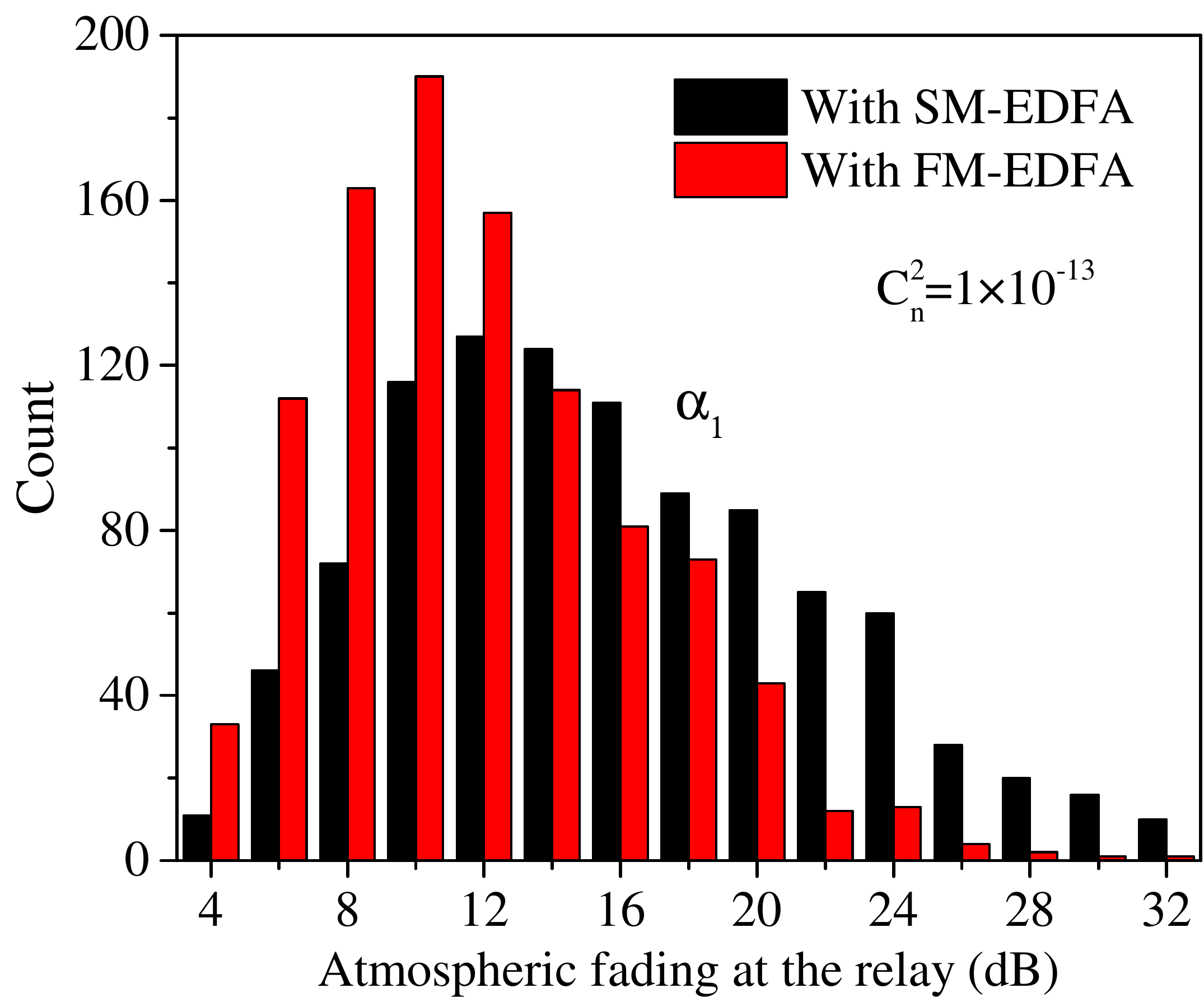}
\caption{The statistical number of $1000$ values of $\alpha_1$ (atmospheric fading of the first hop) in the range of $[2i, 2i+2]$ dB ($i=0, 1, 2, 3, \cdots$).}
\end{figure}

\begin{figure}[htb]
\centering
\includegraphics[width=7.5cm]{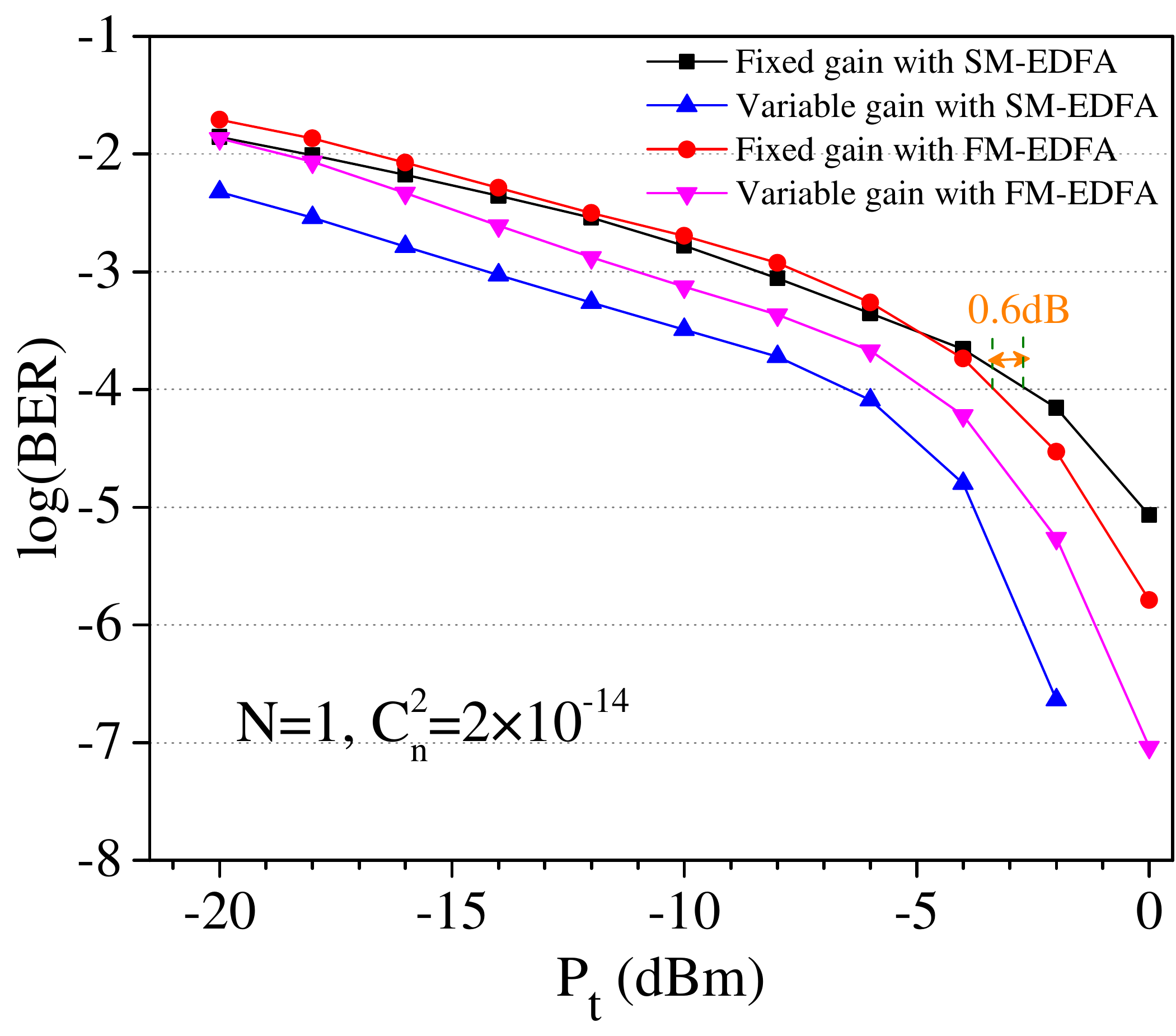}
\caption{BER vs transmitting power under atmospheric turbulence and clear air ($C_n^2=2\times 10^{-14}$, $\alpha_{attn}=0.43dB/km$) with SM-EDFA or with FM-EDFA when a FMF supporting one mode is utilized as receiver at the destination.}
\end{figure}

\begin{figure}[htb]
\centering
\includegraphics[width=7.5cm]{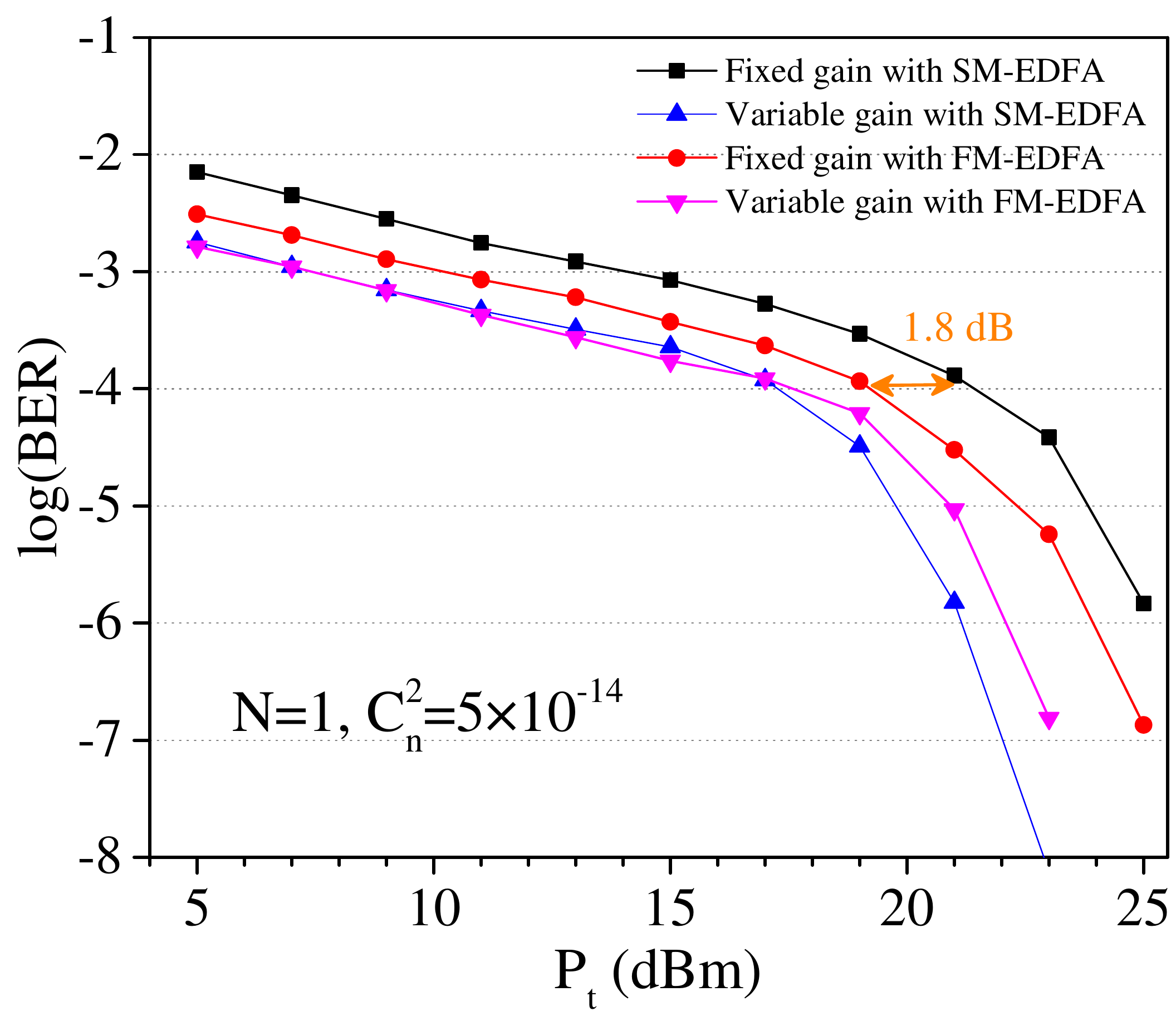}
\caption{BER vs transmitting power under turbulence and haze ($C_n^2=5\times 10^{-14}$, $\alpha_{attn}=4.2dB/km$) with SM-EDFA or with FM-EDFA when a FMF supporting one mode is utilized as receiver at the destination. The performance of both fixed amplification gain and variable amplification gain is considered.}
\end{figure}

\begin{figure}[htb]
\centering
\includegraphics[width=7.5cm]{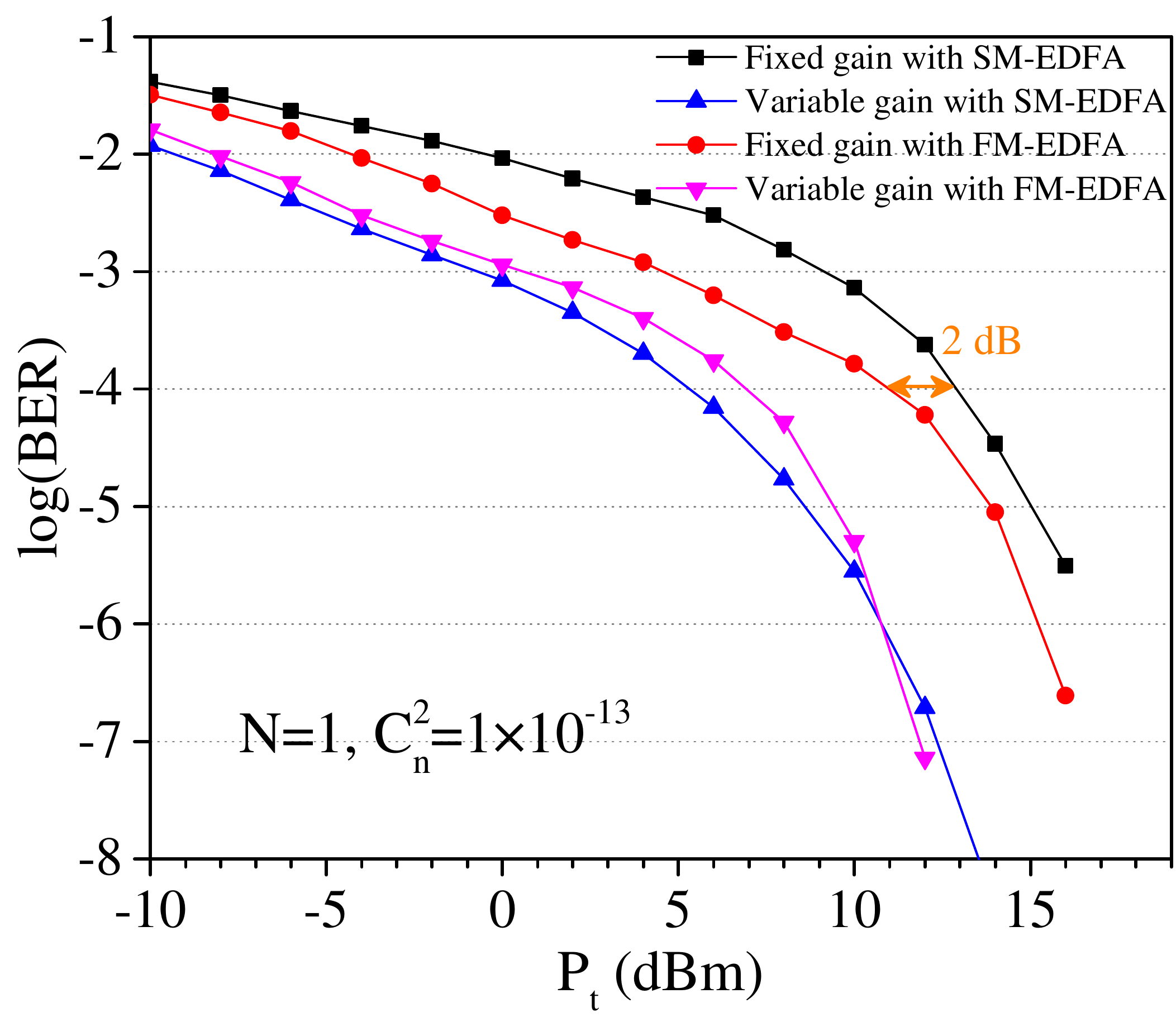}
\caption{BER vs transmitting power under the parameters $C_n^2=1\times 10^{-13}$ and $\alpha_{attn}=0.43dB/km$ for the fixed amplification gain and variable amplification gain relaying system with SM-EDFA or with FM-EDFA when a FMF supporting one mode is utilized as receiver at the destination.}
\end{figure}

\begin{figure}[htb]
\centering
\includegraphics[width=7.5cm]{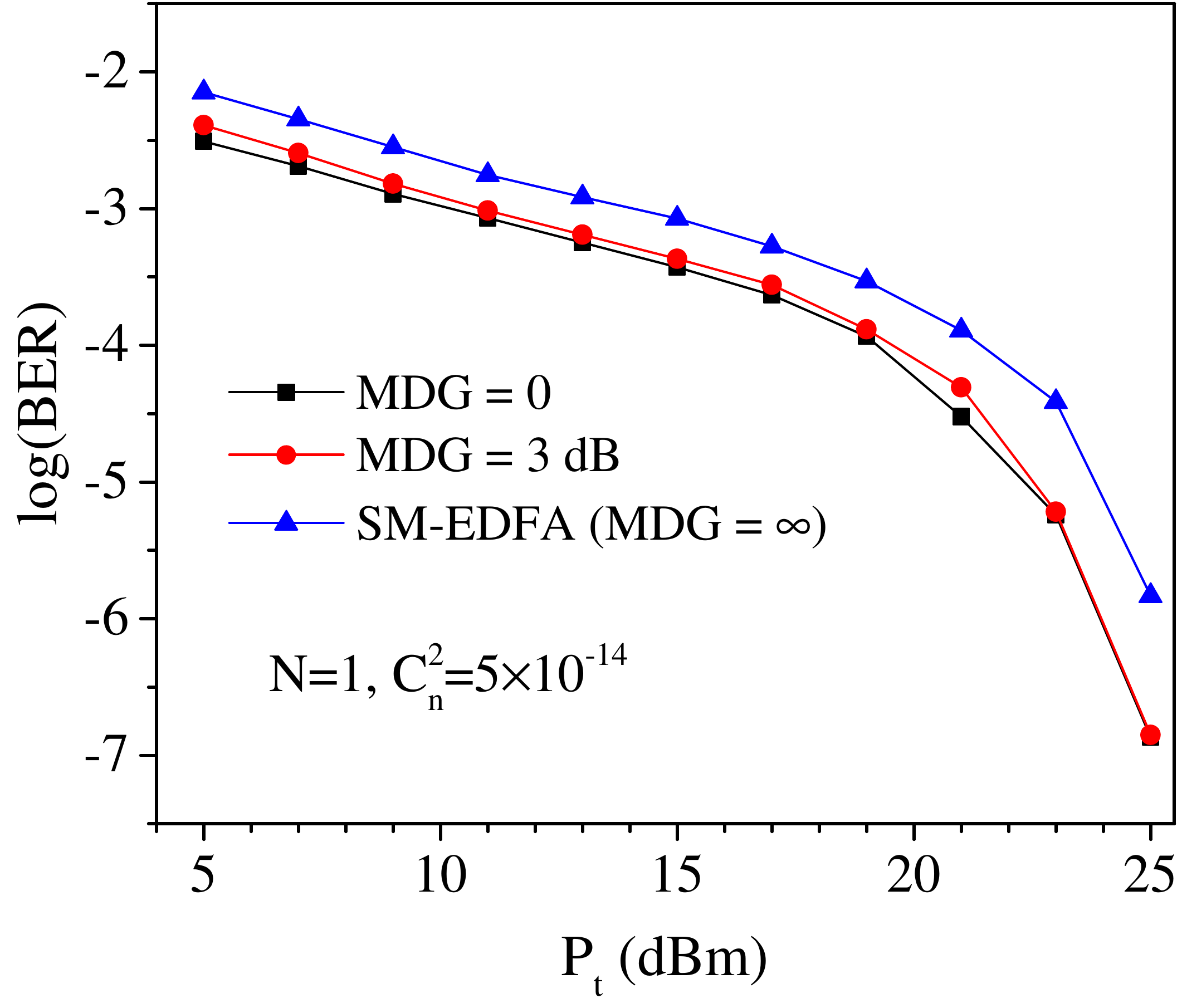}
\caption{BER vs transmitting power for FM-EDFA with different mode-dependent gain (MDG) under weak turbulence and haze ($C_n^2=5\times 10^{-14}$, $\alpha_{attn}=4.2dB/km$) in fixed gain amplification system.}
\end{figure}

\begin{figure}[htb]
\centering
\includegraphics[width=7.5cm]{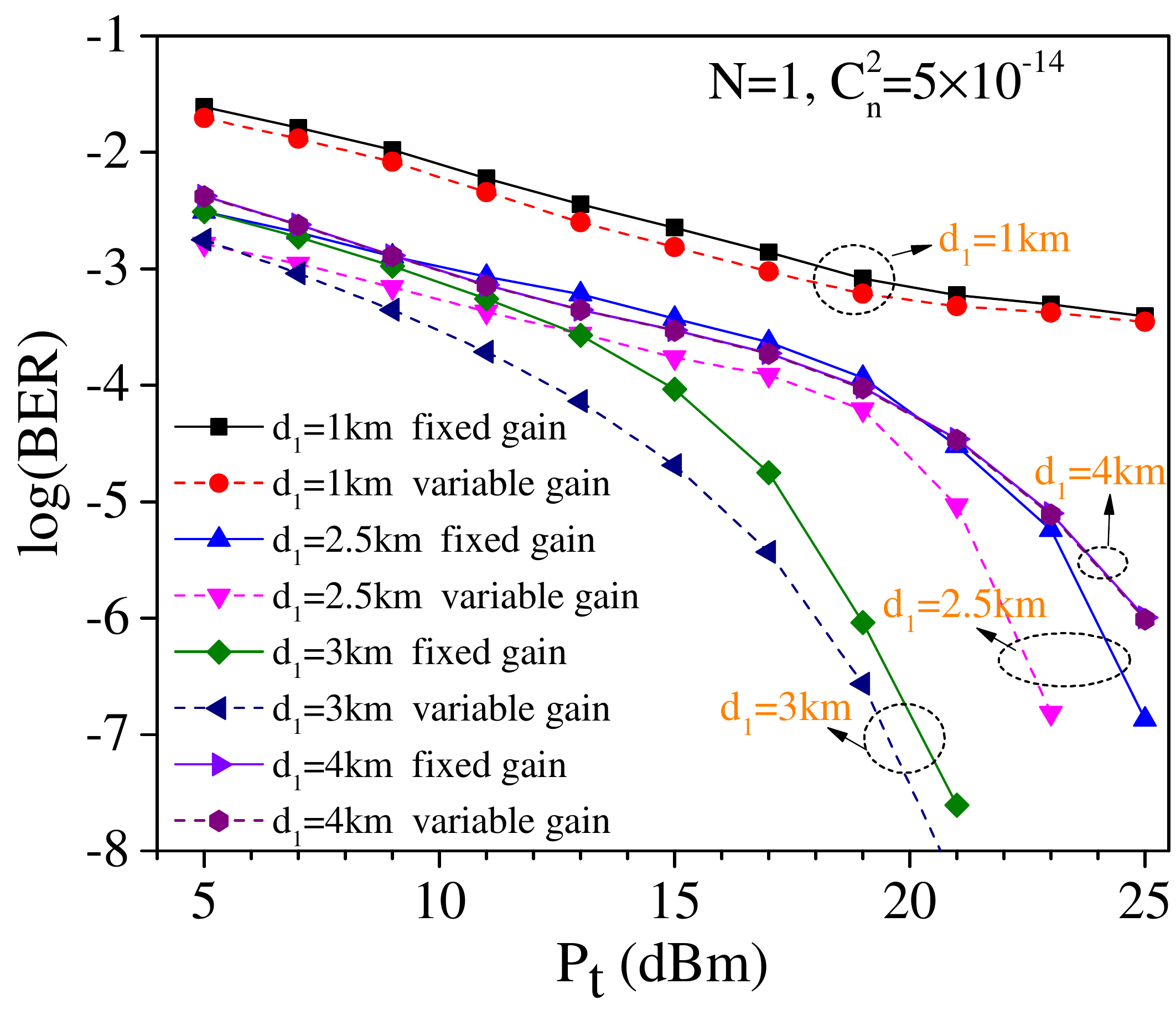}
\caption{BER vs transmitting power for FM-EDFA based relaying system with different relay locations ($d_1 = 1 km$, $2.5 km$, $3 km$ and $4 km$ respectively) under weak turbulence and haze ($C_n^2=5\times 10^{-14}$, $\alpha_{attn}=4.2dB/km$).}
\end{figure}

A FMF which supports N modes (N=1, 2 or 4) is placed on the destination. For $N=1$, the BER for dual-hop relaying system with SM-EDFA or with FM-EDFA is given in Fig. 4-6. Fig. 4 shows the BER vs transmitting power ($P_t$) under fixed and variable gain with $C_n^2=2\times 10^{-14}$ ($D/r_0 = 1.093$) and $\alpha_{attn}=0.43dB/km$. With the increase of transmitting power, the BER of fixed gain relaying system with FM-EDFA decreases faster than that of relaying system with SM-EDFA. The power budget is increased by $0.6 dB$ at $BER = 1\times 10^{-4}$ compared with fixed gain relaying system with SM-EDFA. The reason is mainly due to the increase of signal-dependent noise. When signal-dependent noise is the major noise, the SNR of two relaying system is optimized. At this time, for the turbulence-resistant effect of FM-EDFA, BER performance of fixed gain relaying system with FM-EDFA is better.

With the increase of turbulence level, the growth of power budget for fixed gain relaying system with FM-EDFA is more obvious compared to the fixed gain relaying system with SM-EDFA. Fig. 5 shows the BER performance considering the refractive index structure constant $C_n^2=5\times 10^{-14}$ ($D/r_0 = 1.895$) and attenuation coefficient $\alpha_{attn}=4.2 dB/km$. Obviously, the BER of relaying system with FM-EDFA is always better. The power budget is increased by $1.8 dB$ at $BER = 1\times 10^{-4}$. Furthermore, the power budget is increased by $2 dB$ at $BER = 1\times 10^{-4}$ under $C_n^2=1\times 10^{-13}$ ($D/r_0 = 2.872$) and $\alpha_{attn}=0.43 dB/km$ (Fig. 6). From Fig. 4 to Fig. 6, with the increasing intensity of turbulence, the turbulence compensation effect of FM-EDFA is becoming more and more obviously. In general, the stronger the turbulence is, the better the performance of FM-EDFA based relaying system is than the relaying system with SM-EDFA.

Fig. 4-6 also depict the BER of variable gain relaying system with SM-EDFA or with FM-EDFA as a function of transmitting power. The BER performance of the variable amplification gain is always outperform that of the fixed amplification gain. Different from the fixed gain system, the BER of variable gain relaying system with SM-EDFA is superior to that of variable gain relaying system with FM-EDFA under weak to moderate turbulence ($D/r_0 = 1.093$, $D/r_0 = 1.895$ and $D/r_0 = 2.872$)(Fig. 4-6).

To comment on whether the mode-dependent gain (MDG) of FM-EDFA have an impact on the system performance, MDG of FM-EDFA is taken into account for fixed gain amplification relaying system in Fig. 7. The performance is the best when $MDG = 0$ and a little bit worse when $MDG = 3 dB$ (the gain coefficient of LP$_{\text{01}}$ mode is twice that of LP$_{\text{11}}$ mode).

\begin{figure*}[htb]
\centering
\subfigure{\includegraphics[width=5.9cm]{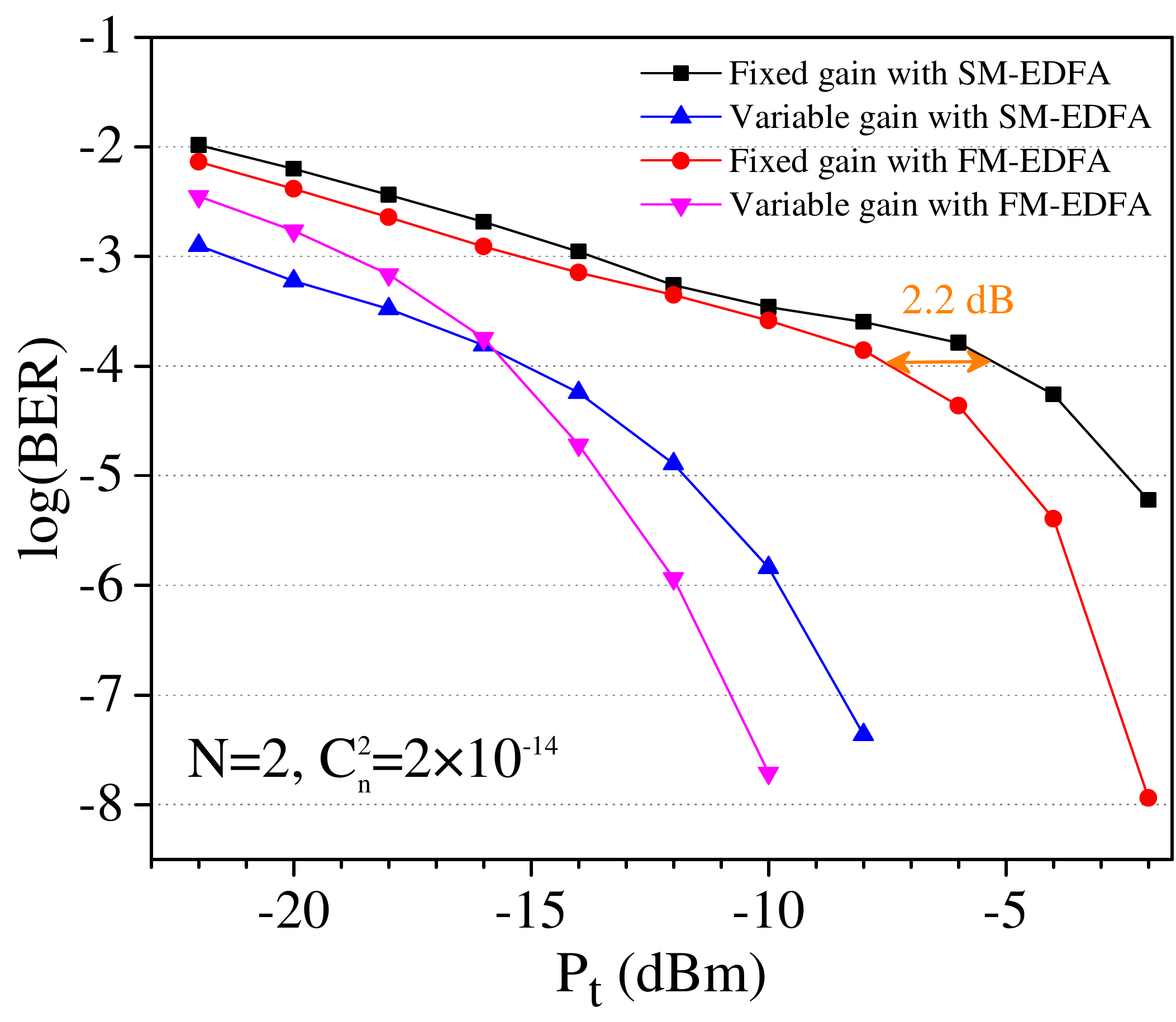}}
\subfigure{\includegraphics[width=5.9cm]{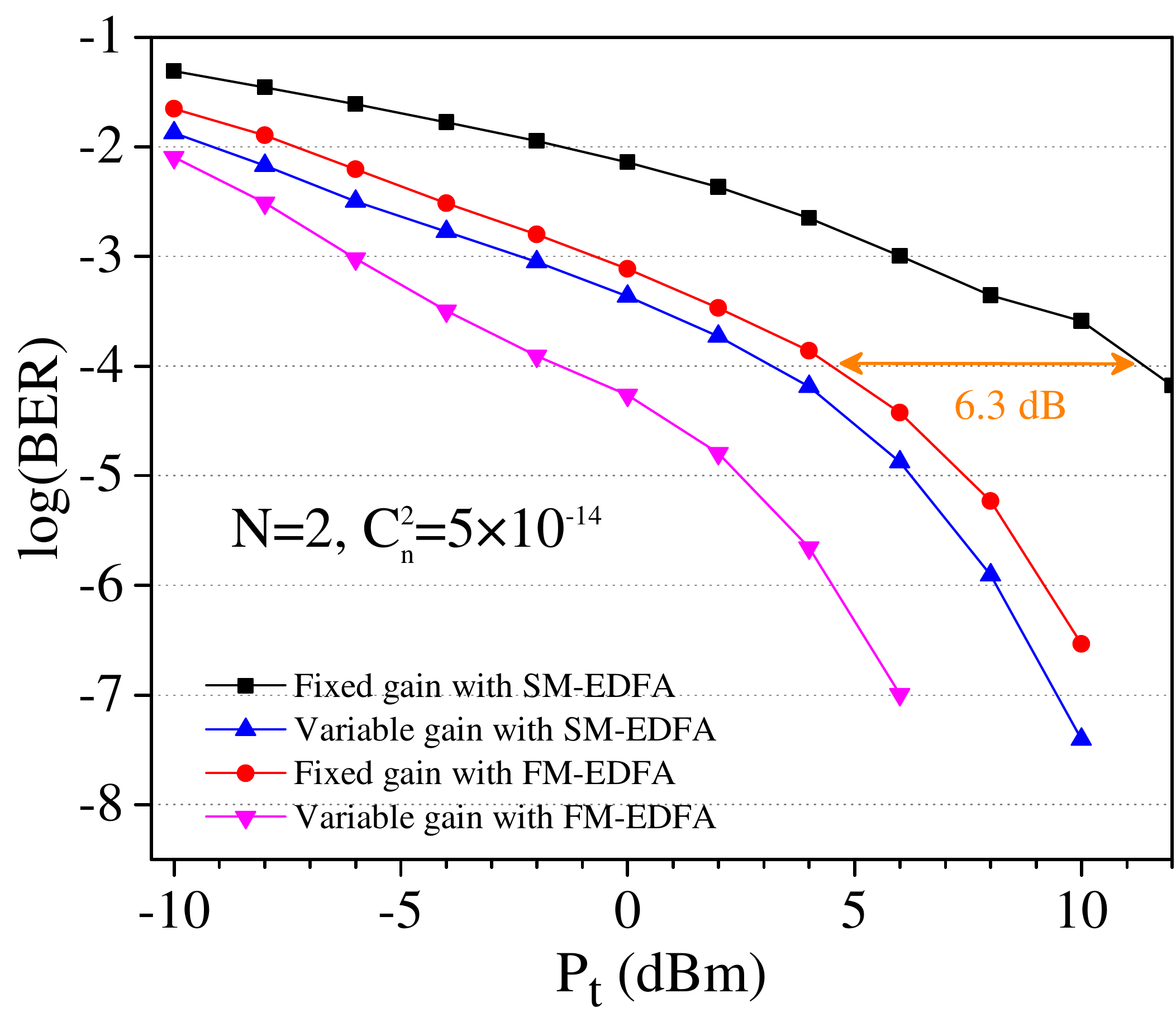}}
\subfigure{\includegraphics[width=5.9cm]{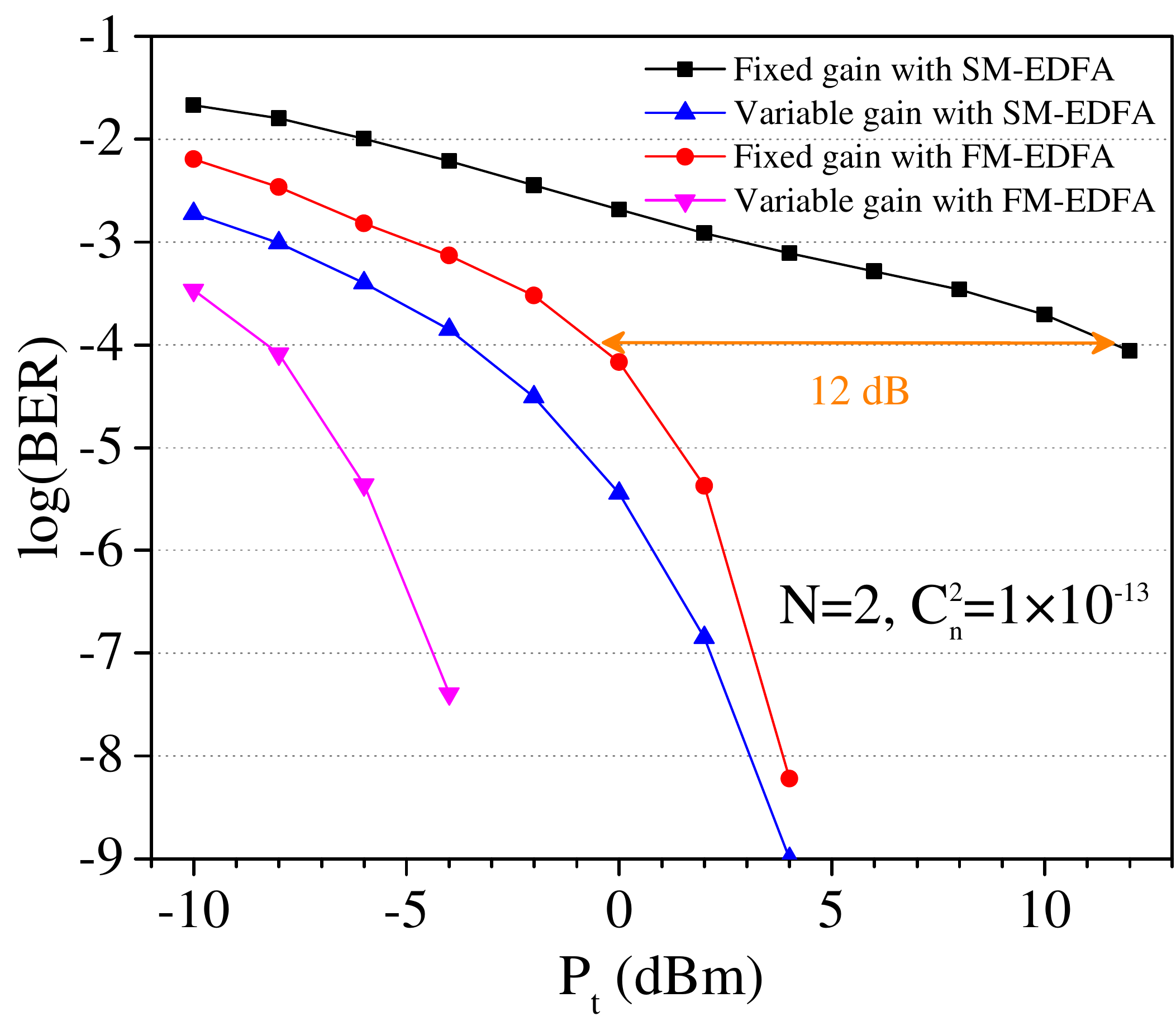}}
\caption{BER vs transmitting power under (left) weak turbulence and clear air ($C_n^2=2\times 10^{-14}$, $\alpha_{attn}=0.43dB/km$), (centre) weak turbulence and haze ($C_n^2=5\times 10^{-14}$, $\alpha_{attn}=4.2dB/km$) and (right) moderate turbulence and clear air ($C_n^2=1\times 10^{-13}$ and $\alpha_{attn}=0.43dB/km$) for the fixed and variable gain amplification relaying system with SM-EDFA or with FM-EDFA. A TMF ($N=2$) is placed on the destination to collect light.}
\end{figure*}

\begin{figure*}[htb]
\centering
\subfigure{\includegraphics[width=5.9cm]{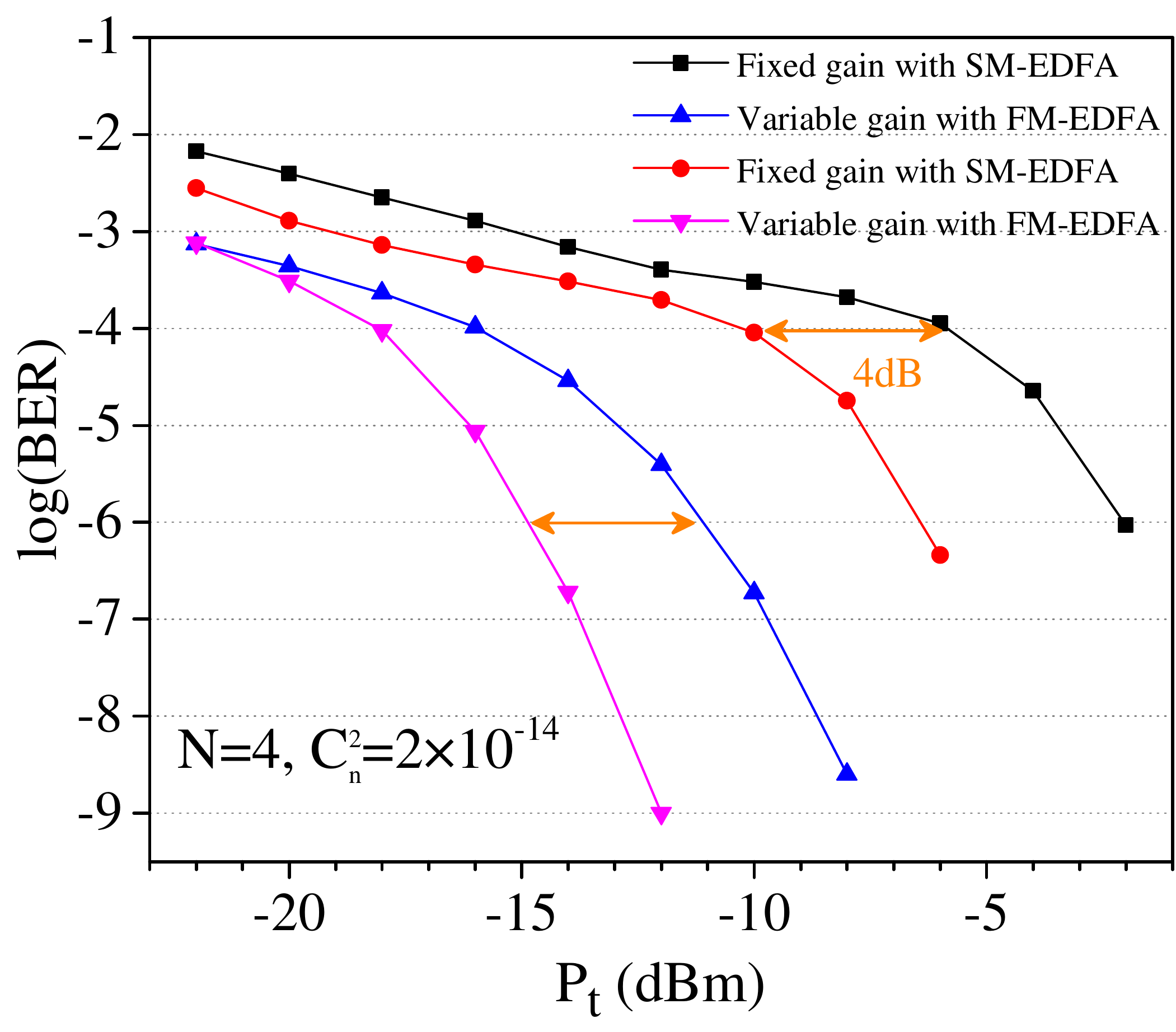}}
\subfigure{\includegraphics[width=5.9cm]{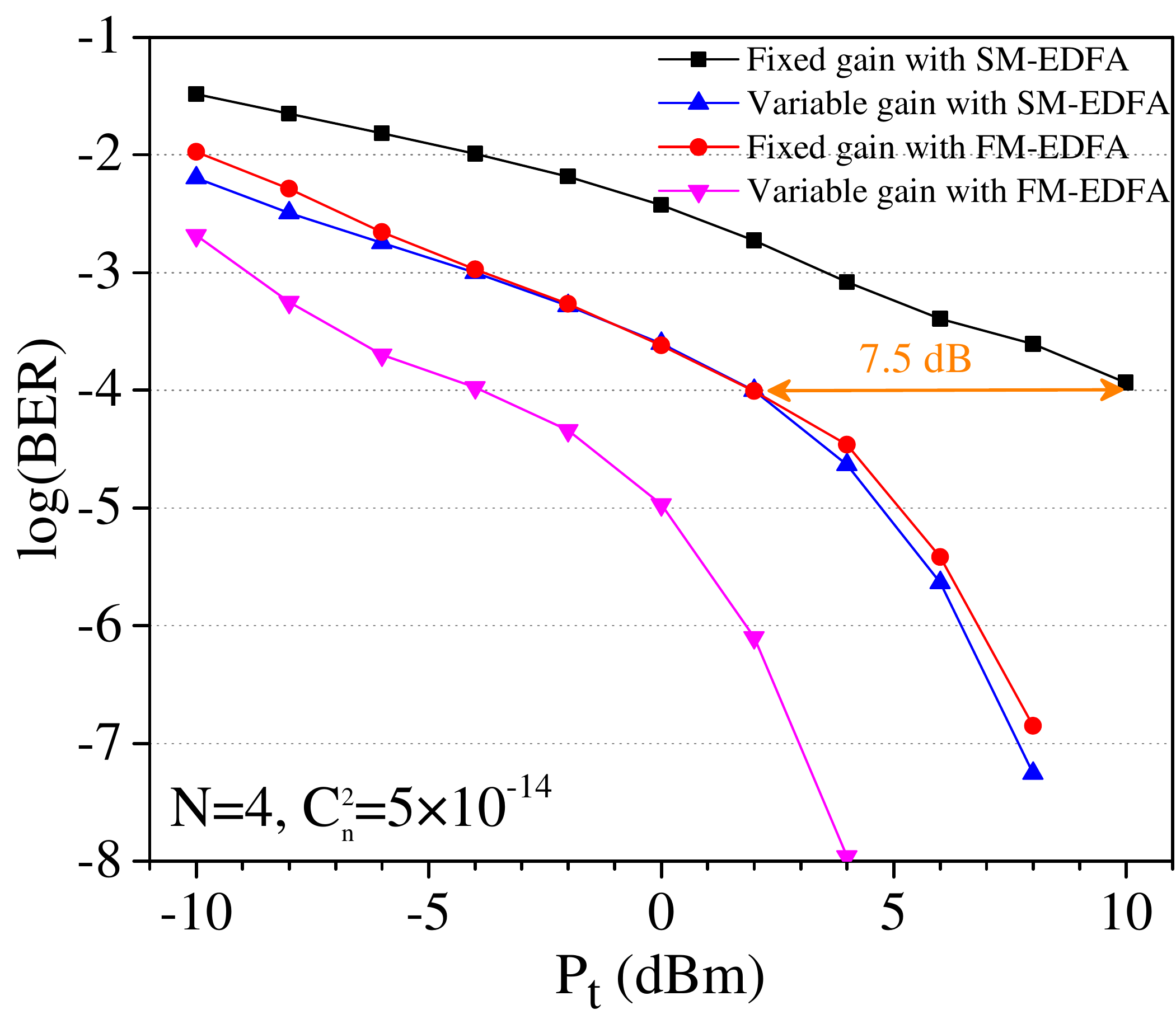}}
\subfigure{\includegraphics[width=5.9cm]{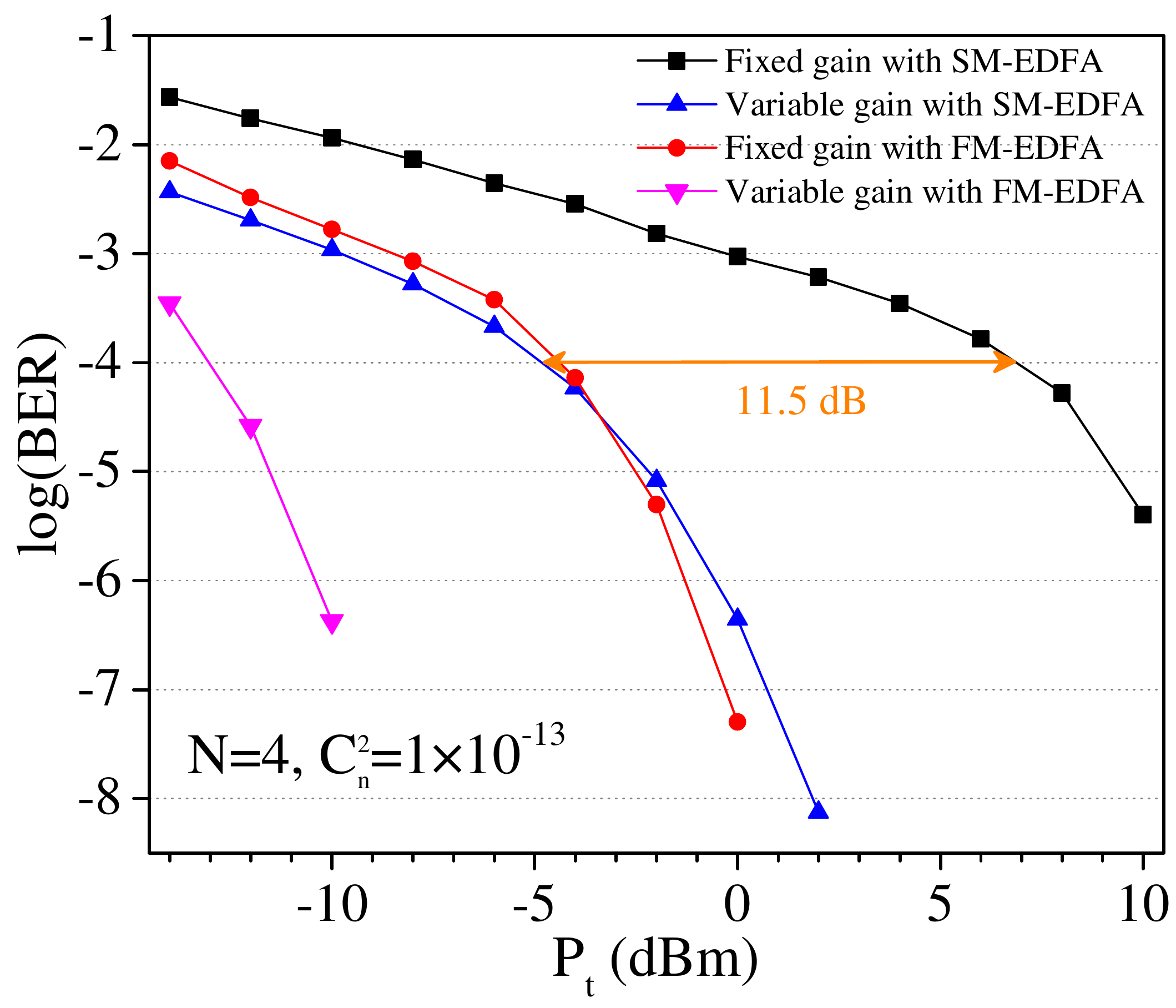}}
\caption{BER vs transmitting power under (left) weak turbulence and clear air ($C_n^2=2\times 10^{-14}$, $\alpha_{attn}=0.43dB/km$) (centre) weak turbulence and haze ($C_n^2=5\times 10^{-14}$, $\alpha_{attn}=4.2dB/km$) and (right) moderate turbulence and clear air ($C_n^2=1\times 10^{-13}$ and $\alpha_{attn}=0.43dB/km$)  with SM-EDFA or with FM-EDFA. A four mode fiber ($N=4$) is placed on the destination.}
\end{figure*}

To analyze the optimal relay placement of dual-hop relaying system with FM-EDFA. Fig. 8 shows the BER performance corresponding to different relay locations $d_1 = 1 km$, $2.5 km$, $3 km$ and $4 km$. $d_1$ is the link lengths of the first hop.  The total source-destination length is $5 km$ ($d_1 + d_2 = 5 km$). The best BER performance occurs at $d_1 = 3 km$ under atmospheric turbulence and haze ($C_n^2=5\times 10^{-14}$, $\alpha_{attn}=4.2dB/km$) when N=1. For fair comparison, the focal lengths of receiving lenses of relaying system with different relay locations are optimized to achieve the maximum coupling efficiency.

Using FMF supporting multiple spatial modes at the destination can further increase the system performance. Fig. 9 and Fig. 10 give the BER for dual-hop relaying system for $N=2$ and $N=4$, respectively. Similar to the relaying system with $N=1$, the BER performance of the fixed gain relaying system with FM-EDFA is superior to that of the fixed gain relaying system with SM-EDFA. For $N=2$, the power budget is increased by $2.2 dB$, $6.3 dB$ and $12 dB$ at $BER = 1\times 10^{-4}$ under the refractive index structure constant $C_{n}^{2} = 2\times 10^{-14}$, $5\times 10^{-14}$ and $1\times 10^{-13}$, respectively. For $N=4$, the power budget is increased by $4 dB$, $7.5 dB$ and $11.5 dB$ at $BER = 1\times 10^{-4}$ under the refractive index structure constant $C_{n}^{2} = 2\times 10^{-14}$, $5\times 10^{-14}$ and $1\times 10^{-13}$, respectively. Under the same $C_n^2$, at $BER = 1\times 10^{-4}$, the increase of power budget for FM-EDFA based relaying systems with $N=2$ or $N=4$ are larger than that with $N=1$. Moreover, when a four mode fiber is placed on the destination, the BER of the variable gain system with FM-EDFA is also always better than that of the variable gain system with SM-EDFA (Fig. 10) which is different from the relaying system with $N=1$ (Fig. 4-6). In addition, under the same turbulence level, the bigger the N-value, the smaller the transmitting power required for the same BER.

\section{Conclusion}
A FM-EDFA based all-optical relaying system is proposed in this paper to resist atmospheric turbulence. Atmospheric turbulence is modeled with turbulence phase screen. At the relay, using FM-EDFA as the receiver leads to the improvement of the turbulence induced fading which is expected to bring improved BER for all optical relaying system. A model for dual-hop FSO system with FM-EDFA is derived and the BER performance of FM-EDFA based relaying system is given. Compared with relaying system based on SM-EDFA, the BER of FM-EDFA based relaying system is always superior to that of SM-EDFA based relaying system for the fixed gain amplification. When a FMF supporting 2 or 4 spatial modes is placed at the destination, the BER performance of relaying system will be further improved. Under moderate turbulence $D/r_0 = 2.872$ ($C_{n}^{2} = 1\times 10^{-13}$), at $BER = 1\times 10^{-4}$, the power budget of the fixed gain relaying system with FM-EDFA is increased more than $10 dB$ than that of the fixed gain relaying system with SM-EDFA. In general, the proposed FM-EDFA based all-optical relaying systems have better communication capacity for its turbulence suppression ability and have potential to be used in future last-mile FSO systems.


%

\ifCLASSOPTIONcaptionsoff
  \newpage
\fi



%

%


\begin{IEEEbiographynophoto}{Shanyong Cai}
received his Ph.D. degree in information and communication engineering from Beijing University of Posts and Telecommunications (BUPT), Beijing, China, in 2017 and the M.S. degree in optics from South China Normal University, Guangzhou, China, in 2013. He currently holds a postdoctoral position with the Institute of Information Photonics and Optical Communications, BUPT. His research interests include space-division multiplexing (SDM), free space optical communication (FSO) and orthogonal frequency-division multiple access (OFDMA) PON.
\end{IEEEbiographynophoto}

\begin{IEEEbiographynophoto}{Zhiguo Zhang}
received the B.S. degree from Shandong University, Shandong, China, in 2002 and the Ph.D. degree from Beijing University of Posts and Telecommunications (BUPT), Beijing, China, in 2007. He is now an associate professor of Institute of Information Photonics and Optical Communications, BUPT. His main research interests include free space optical communication, optical fiber sensor and optical access networks.
\end{IEEEbiographynophoto}

\begin{IEEEbiographynophoto}{Xue Chen}
received the B.S. degree from Dalian University of Technology, Dalian, China, in 1982 and the M.S. degree from Beijing University of Posts and Telecommunications (BUPT), Beijing, China, in 1985. She is now a professor of Institute of Information Photonics and Optical Communications, BUPT. Her main research interests focus on backbone optical transmission and optical access networks.
\end{IEEEbiographynophoto}




\end{document}